# Electrically tunable moiré magnetism in twisted double bilayers of chromium triiodide

Guanghui Cheng[1,2,3,4,12], Mohammad Mushfiqur Rahman[5,12], Andres Llacsahuanga Allcca[1,2,6], Avinash Rustagi[5], Xingtao Liu[1,7], Lina Liu[1,2], Lei Fu[1,2], Yanglin Zhu[8], Zhiqiang Mao[8], Kenji Watanabe[9], Takashi Taniguchi[10], Pramey Upadhyaya[2,5,6]*, Yong P. Chen[1,2,4,5,6,11]*

[1]Department of Physics and Astronomy, and Birck Nanotechnology Center, Purdue University; West Lafayette, Indiana 47907, USA.

[2]Purdue Quantum Science and Engineering Institute, Purdue University; West Lafayette, Indiana 47907, USA.

[3]Department of Physics, University of Science and Technology of China; Hefei, Anhui 230026, China.

[4]WPI-AIMR International Research Center for Materials Sciences, Tohoku University; Sendai 980-8577, Japan.

[5]School of Electrical and Computer Engineering, Purdue University; West Lafayette, Indiana 47907, USA.

[6]Quantum Science Center; Oak Ridge, Tennessee 37831 USA.

[7]School of Industrial Engineering, Purdue University, West Lafayette, Indiana 47907, USA.

[8]Department of Physics, Pennsylvania State University; University Park, Pennsylvania 16802, USA.

[9]Research Center for Functional Materials, National Institute for Materials Science; 1-1 Namiki, Tsukuba 305-0044, Japan.

[10]International Center for Materials Nanoarchitectonics, National Institute for Materials Science; 1-1 Namiki, Tsukuba 305-0044, Japan.

[11]Institute of Physics and Astronomy and Villum Centers for Dirac Materials and for Hybrid Quantum Materials, Aarhus University; 8000 Aarhus-C, Denmark.

[12]These authors contributed equally: Guanghui Cheng, Mohammad Mushfiqur Rahman.

*Corresponding authors. Emails: prameyup@purdue.edu; yongchen@purdue.edu

**Moiré superlattices in van der Waals structures can be used to control the electronic properties of the material and lead to emergent correlated and topological phenomena. Its first demonstration in van der Waals magnets exhibited noncollinear states and domain structures with, however, limited manipulation. Here we report electrically tunable moiré magnetism in twisted double bilayers — that is, a bilayer plus a bilayer with a twist angle between them — of layered antiferromagnet $CrI_3$. Using magneto-optical Kerr effect microscopy, we observe the coexistence of antiferromagnetic and ferromagnetic order with nonzero net magnetization — a hallmark of moiré magnetism. Such magnetic state extends over a wide range of twist angles (with transitions at around 0° and above 20°) and exhibits a nonmonotonic temperature dependence. We further demonstrate voltage-assisted magnetic switching. The observed nontrivial magnetic states and unprecedented control by twist angle, temperature and electrical gating are supported by the simulated phase diagram of the moiré magnetism.**

## Introduction

Moiré superlattices are created by stacking one layer of a van der Waals (vdW) material onto another layer with either a lattice mismatch or a small twist angle[1-13]. Such large superperiodic patterns define a new crystal structure with electronic bands folded into a mini-Brillouin zone, which can form flat bands and can drive the system into strongly correlated regimes. Due to the control they offer over the charge degree of freedom, moiré superlattices are a versatile electronic quantum simulator and platform to discover new phases of matter. A range of properties — including superconductivity, Mott insulating states, and moiré excitons — have been observed in twisted graphene, transition metal dichalcogenides, and other electronic materials[1-6, 8, 13]. It is highly desired to further expand such explorations to additional material types and (beyond charge) degrees of freedom.

VdW magnets exhibit two-dimensional magnetic phenomena and can be used to manipulate the spin degree of freedom [14-20]. Theoretical analyses have predicted that twisted bilayers of vdW magnets can host exotic magnetic phases [21-24] suggesting, in particular, that noncollinear spin textures, topological skyrmion lattices, and rich spectra of magnonic subbands can emerge from the modulation of interlayer exchange interactions [21-24].

The layered antiferromagnet chromium triiodide ($CrI_3$) is of potential interest in this regard. The interlayer magnetic exchange of bilayer $CrI_3$ depends on the structural stacking order[18, 25]: monoclinic stacking favoring antiferromagnetic (AFM) coupling (as in "natural" bilayers, with zero net magnetization) and rhombohedral stacking favoring ferromagnetic (FM) coupling. Twisting one $CrI_3$ layer relative to another layer can create a moiré superlattice with rhombus-shaped primitive cells (Fig. 1a). The atomic registry varies continuously, leading to alternating structural domains such as AA stacking (top and bottom Cr atoms are aligned vertically), rhombohedral stacking, and monoclinic stacking. Therefore, the stacking-dependent interlayer exchange in twisted $CrI_3$ can give rise to magnetic domains and noncollinear spin textures.

Such coexisting AFM and FM states have recently been experimentally demonstrated in twisted $CrI_3$ using magneto-optical measurements and Raman spectroscopy[11, 12], and the nanoscale magnetic domains have been visualized with quantum magnetometry[10]. Twisted vdW magnets as a potential platform to explore and engineer nontrivial magnetic phases. However, effective manipulation of moiré magnetism — including magnetization reversal using electrical methods — remains challenging[17, 26, 27].

In this article, we report electrically tunable moiré magnetism in twisted double bilayer (tDB) $CrI_3$ (that is, a bilayer plus a bilayer with a twist angle between them). We show that antiferromagnetic and ferromagnetic orders can be tuned by the twist angle. We also observe nonmonotonic temperature dependence of the magnetization and demonstrate electrical control over the magnetic states. Our experimental observations are supported by theoretical analysis, and we develop a phase diagram of moiré magnetism in the tDB system, which hosts a rich class of noncollinear states beyond those observed in twisted bilayer (single layer plus single layer) magnets[10, 11] due to the increased layer degree of freedom.

## Moiré superlattice of tDB $CrI_3$

We fabricated tDB $CrI_3$ by the tear-and-stack technique[13, 28] in an argon glovebox and encapsulated it with hexagonal boron nitride (hBN) flakes (Methods). The atomic force microscopy image (Supplementary Fig. 1) suggests a uniform stacking interface with few bubbles. We employ magneto-optical Kerr effect microscopy (MOKE) under polar configuration as the primary measurement due to its high sensitivity to magnetic moments perpendicular to the sample surface[29]. All MOKE measurements were performed under a perpendicular magnetic field at a temperature of 6 K unless otherwise specified (Methods).

We employ high-angle annular dark-field scanning transmission electron microscopy (STEM) to characterize the moiré superlattice of tDB $CrI_3$ (Methods). Figure 1b shows the fast Fourier transform (FFT) of the STEM image of a tDB $CrI_3$ with a target twist angle of 1.42° (see the original STEM images in Supplementary Fig. 2). Two sets of first- and second-order Bragg peaks with sixfold rotation symmetry are marked by the green and red rectangles. A close inspection of one representative peak reveals two slightly separated peaks corresponding to the top and bottom bilayer $CrI_3$. As a comparison, no splitting is observed for the Bragg peaks of the natural bilayer $CrI_3$ (Supplementary Fig. 2). From the splitting, we can measure the actual twist angle to be ~1.9°, within the accuracy of ~±0.5° of the stacking processes. Such a twist-angle accuracy is comparable with that in other works[10-12]. In this paper, unless otherwise specified, all twist angles (denoted $\theta$) of the experimental samples refer to the target twist angles. The moiré wavelength[30] can be estimated to be $L = \frac{a}{2\sin(\theta/2)}$, where $a$ is the lattice constant of $CrI_3$[24] and is ~ 20.5 nm for the above sample. By inverse FFT of the two sets of Bragg peaks marked by green and red rectangles in Fig. 1b, the real-space moiré patterns are reconstructed in Figs. 1c,d, corresponding to the first and second orders of the moiré superlattice. Both pattern periods are consistent with the expectations $L$ and $L/\sqrt{3}$ based on the moiré wavelength. The STEM characterizations demonstrate the formation of a moiré superlattice in tDB $CrI_3$.

**Twist angle and temperature dependence of the magnetic behaviors**

Figure 2a shows the optical micrograph of a representative bilayer $CrI_3$ flake (left) and a 5°-tDB $CrI_3$ (right) made from it. The MOKE signals as a function of magnetic field in bilayer $CrI_3$ and 5°-tDB $CrI_3$ are shown in Figs. 2b,c, measured at the yellow and red spots in Fig. 2a, respectively. In the bilayer $CrI_3$ region, we observed zero magnetization at zero field and sharp jumps at fields of ±0.7 T, consistent with the reported AFM spin-flip transition in natural bilayer $CrI_3$[31, 32]. The magnetic ground states are shown in the insets.

In stark contrast, within the tDB $CrI_3$ region, in addition to the AFM spin-flip transitions present at fields of ~ ±0.7 T (since these are analogous to the transitions observed in bilayer $CrI_3$, we refer to them hereafter as "AFM spin-flip transitions" even for the tDB system), a new and significant *FM hysteresis loop* emerges with transition fields of ~ ±0.2 T. In comparison, the MOKE results of natural four-layer $CrI_3$ suggest interlayer AFM coupling without an FM loop[32] (Supplementary Fig. 3). For tDB $CrI_3$, the assumption of interlayer AFM spin orientations within each of the top and bottom bilayers breaks down, since it leads to zero net magnetization regardless of the coupling type between the middle two layers. Here, we propose that the observed AFM and FM behaviors in tDB $CrI_3$ are due to the coexistence of AFM/FM domains predicted for moiré magnetism[21, 23]. While similar coexistence of magnetic domains has been reported in twisted bilayer $CrI_3$[10-12], our tDBs could have even richer magnetic phase diagram and behaviors, as discussed later.

The moiré magnetism is essentially determined by competing magnetic interactions dependent on the twist angle[21, 24]. We thus studied the twist angle dependence of the magnetic behaviors in tDB $CrI_3$. Figures 2d,e show the MOKE results in 0°- and 30°-tDB $CrI_3$, respectively (0°-tDB $CrI_3$ refers to stacking bilayer on bilayer with 0°, instead of a natural four-layer $CrI_3$). Only AFM spin-flip transitions are present, while the FM loop disappears, in sharp contrast to 5°-tDB $CrI_3$. Data from other samples with additional twist angles are shown in Supplementary Fig. 4. Furthermore, MOKE results of a representative tDB $CrI_3$ sample exhibiting both FM loop and AFM spin-flip transitions show no other transitions up to higher fields of ±2 T (Supplementary Fig. 5).

We denote the magnitudes of the FM loop and AFM spin-flip transition as $\Delta\theta_{K1}$, $\Delta\theta_{K2}$ and summarize the phenomenological "FM fraction" $\Delta\theta_{K1}/(\Delta\theta_{K1}+\Delta\theta_{K2})$ as a function of the twist angle in Fig. 2f. The data points at each twist angle are obtained at different positions in each sample (see uniformity check in Supplementary Fig. 6). The nonzero fractions (reaching a maximum value of ~0.3) in the twist angle range of 0.14° - 20° indicate a regime with coexisting FM and AFM behaviors. The FM loop disappears for twist angles of ~0° or above 20°, suggesting phase transitions across critical angles. Such systematic twist angle dependence and reproducible results support the noncollinear moiré magnetism emerging in tDB $CrI_3$. The possible noncollinear phases will be discussed with the theoretically calculated phase diagram shown later. We further plot another quantity, the transition fields of both FM coercivity and AFM spin-flip, as a function of the twist angle in Supplementary Fig. 7. No significant dependence is observed, suggesting that the magnetic anisotropy in this system is hardly tuned by the moiré superlattice. We note that compared with other recent measurements in tDB $CrI_3$ system [12, 33], the magnetic transitions look sharper in our MOKE curves, possibly related to the background subtraction in our MOKE signal and disorder-dependent coercivity [34].

We further investigate the temperature dependence of the magnetic behaviors. Figures 3a,b show the MOKE results for bilayer $CrI_3$ and 0.71°-tDB $CrI_3$ at various temperatures. With increasing temperature, the FM coercivity and AFM spin-flip transitions occur at lower fields and with reduced MOKE amplitudes (except for the anomalous nonmonotonic behavior of the FM loop amplitude for the tDB $CrI_3$, to be discussed later), eventually disappearing at 40 ~ 50 K, consistent with the studies in natural bilayer $CrI_3$ and twisted bilayer $CrI_3$ [11, 17, 31]. Careful inspection reveals that the critical temperature of the FM coercivity is somewhat higher than that of the AFM spin-flip transition, possibly because the two transitions (AFM vs FM) correspond to spin reorientation involving different regions and different types of couplings of the tDB system (see Supplementary Figs. 13-16 for the expected spin reorientations corresponding to these transitions); consequently, the coercivities and the net change in magnetization corresponding to these transitions are controlled by different material parameters, possibly giving rise to different critical temperatures.

Figure 3c shows the MOKE magnitudes of the AFM spin-flip transition $\Delta\theta_{K0}$ in the natural bilayer, and the FM loop $\Delta\theta_{K1}$, the AFM spin-flip transition $\Delta\theta_{K2}$ and the total magnitude $\Delta\theta_{K1} + \Delta\theta_{K2}$ in tDB as functions of temperature extracted from Figs. 3a,b. Remarkably, $\Delta\theta_{K1}$ of 0.71°-tDB $CrI_3$ (proportional to the net out-of-plane magnetization within the noncollinear phase) first increases and then drops down with increasing temperature. In contrast, $\Delta\theta_{K0}$ of bilayer $CrI_3$ and $\Delta\theta_{K2}$, $\Delta\theta_{K1}+\Delta\theta_{K2}$ of 0.71°-tDB $CrI_3$ (proportional to the saturation magnetization of the sample, $M_s$) monotonically decrease with increasing temperature, typical of magnets making a transition from a ferromagnetic phase to a paramagnetic phase. Supplementary Fig. 8 presents additional examples of such a nonmonotonic temperature dependence of $\Delta\theta_{K1}$ observed in another tDB $CrI_3$ sample. As discussed below, such anomalous nonmonotonic behavior of $\Delta\theta_{K1}$ is further consistent with the presence and behaviors of noncollinear phases predicted in this system. Namely, it can be qualitatively understood as the temperature-induced crossover between different noncollinear phases arising from different power-law dependences of the magnetic parameters (anisotropy $K$, interlayer exchange $J_{\perp}^{0}$ and intralayer exchange $A$) on the temperature[11, 35, 36].

### Electrical control of the magnetic behaviors

We next explore the electrical control of the magnetic behaviors in tDB $CrI_3$. We fabricated a back-gated 0.14°-tDB $CrI_3$ device (Methods), as shown in Figs. 4a,b. The MOKE results at a gate voltage of -80 V (Fig. 4c) present the coexistence of AFM and FM transitions, the same feature as discussed above for noncollinear moiré magnetism. Four plateaus at low fields representing distinct magnetic states are denoted by A, B, C and D (presumably with positive, negative, negative and positive magnetization, respectively, with the MOKE signal sign consistent with the correspondingly polarized state at high fields). Figure 4d shows the MOKE signal as a function of the magnetic field at various back-gate voltages (see representative MOKE curves in Supplementary Fig. 9). The top and bottom panels correspond to backward and forward sweeps of the magnetic field, respectively. The sharp color boundaries represent FM coercivity and AFM spin-flip transitions. Both occur at lower magnetic fields with increasing back-gate voltage, suggesting a high electrical tunability of the magnetic anisotropy and interlayer exchange. Similar electric tunability is observed in natural bilayer $CrI_3$[17, 27, 37], which is associated with doping-modulated interlayer exchange coupling.

Remarkably, we observed voltage-controlled magnetic switching in tDB $CrI_3$. Such switching is essentially due to the electrical modulation of coercive fields. Here, we biased the magnetic field close to the FM coercive field denoted by the circled numbers in Fig. 4c. The corresponding MOKE signal as a function of back-gate voltage $V_{bg}$ is shown in Fig. 4e. At a fixed voltage $V_{bg}$ = -80 V, the tDB $CrI_3$ sample is first initialized by applying a high magnetic field of 1 T and then biased at a field of -0.141 T (trace 1), corresponding to a noncollinear ground state (A state, with positive magnetization). When sweeping $V_{bg}$ from -80 V to 80 V, an abrupt switching to the B state occurs at ~70 V. For biased fields of -0.151 T (trace 2) and -0.16 T (trace 3), which are still higher but closer to the coercive field (-0.178 T) at $V_{bg}$ = -80 V, switching occurs at earlier voltages of -10 V and -50 V, respectively. This is consistent with the trend of decreasing coercive field magnitude for more positive $V_{bg}$ (Fig. 4d and Fig. S9) for this sample. However, the system stays at the B state when sweeping the voltage backward, presumably because the B state with negative magnetization is the low-energy state under the negative magnetic field, and an energy barrier prevents the system from returning to the A state. Such one-time magnetic switching has also been reported in other ferromagnetic systems[27, 38]. Conversely, for the initialization by a negative high field of -1 T and biased fields of 0.14 T, 0.15 T, 0.153 T (traces 4, 5, 6, increasingly close to the coercive field of 0.174 T at $V_{bg}$ = -80 V for the forward magnetic field sweep), voltage-controlled switching is observed from a noncollinear state with negative magnetization (C state) to that with positive magnetization (D state). The voltage-assisted switching is also observed in other tDB $CrI_3$ samples with twist angles of 5° and 20° (Supplementary Fig. 10).

Besides voltage-assisted switching, we further observed a linear dependence of the MOKE signal on the voltage with a positive slope for traces 1, 2, and 3 in Fig. 4e. This suggests an increase in the net magnetization (before switching occurs) as a function of the gate voltage (interestingly, this is true for the backward gate voltage sweep as well, where magnetization becomes increasingly negative). When the magnetic field/magnetization is reversed (traces 4, 5, 6), the magnetization versus voltage shows the opposite slope, but the net magnetization (before switching) still generally increases during voltage sweeps. Such linear dependences were also measured in 2 other samples (5°- and 20°-tDB, Supplementary Fig. S10), but the sign of the slope was opposite in the latter sample (20°-tDB, with generally decreasing magnetization magnitudes during voltage sweeps, Supplementary Fig. S10f). These behaviors in response to voltage demonstrate intriguing electrical tuning of distinct time-reversal states (A, B states versus C, D states) in tDB $CrI_3$ as a result of different samples and initialization processes. The observed gate-tunable magnetism may have two possible origins. The electrostatic field can break the spatial-inversion symmetry of the layered magnet and create layer polarization of spin carriers [17, 37, 39]. Furthermore, the gate voltage can effectively change the magnetic parameters of $CrI_3$ by electrostatic doping [27, 40], which could change the location of the system on the phase diagram below (Fig. 5) and hence modulate magnetization in twisted bilayers. We notice that different slopes between different samples could arise from different locations and (voltage-swept) trajectories on the parameter space and phase diagram (Supplementary Note 11). The observed electrical tunability allows us to access and control the noncollinear magnetic states in tDB $CrI_3$.

## Theoretical analysis of moiré magnetism: phase diagram

To better understand the observations, we theoretically study the moiré magnetism in tDB $CrI_3$. Within the continuum approximation[21], the free energy per unit area for tDB $CrI_3$ with a relative twist between the middle two layers can be written as:

$$\mathcal{F}(\vec{m}_1,\vec{m}_2,\vec{m}_3,\vec{m}_4) = A\sum_{i=1}^{4}(\nabla\vec{m}_i)^2 - K\sum_{i=1}^{4} m_{iz}^2 + J_{\perp}^{0}[\vec{m}_1\cdot\vec{m}_2 + \Phi(\vec{r})\vec{m}_2\cdot\vec{m}_3 + \vec{m}_3\cdot\vec{m}_4],$$

where $\vec{m}_i$ is the position ($\vec{r}$)-dependent unit vector oriented along the magnetization of layer $i$, $A$ is the intralayer exchange stiffness, $K > 0$ is the easy axis anisotropy, and $J_{\perp}^{0}$ is the interlayer exchange constant for monoclinic stacking. $\Phi(\vec{r})$ accounts for the spatially-dependent interlayer exchange due to local variation of stacking between the twisted layers, acting as a new degree of freedom tunable via the twist angle. The relative strengths of the intralayer, interlayer exchange and anisotropy energies determine the magnetic ground state of the system. Following the framework of Ref. [21], we define two dimensionless parameters $\alpha \equiv \frac{J_{\perp}^{0}}{Aq_m^2}$ and $\beta \equiv \frac{K}{Aq_m^2}$ that capture the relative strengths between these energy contributions, where $q_m=1/L$ is the moiré wavevector.

Owing to the increased "layer degree of freedom" of the tDB system, a richer set of magnetic configurations beyond the twisted bilayer case[10-12, 21, 24] is necessary to characterize the possible ground states of our system as functions of $\alpha$ and $\beta$. We schematically depict them in Fig. 5a (for convenience, we label the layers from bottom to top as 1, 2, 3, 4, and do not show the copies obtained by: (i) time reversal, or (ii) simultaneous swapping of the spin configurations between layers 1↔4 and 2↔3 so that layers 1, 2, 3, 4 are now from top to bottom). When the configurations are governed by minimizing the free energy terms due to perpendicular anisotropy and interlayer exchange, the magnetization within each layer prefers to be primarily out-of-plane and have the interlayer spin orientations which minimize the *local* interlayer exchange (similar to the so-called twisted-A phase for twisted bilayer $CrI_3$ [21]). For tDB $CrI_3$, the resultant configuration harbors ~180° domain walls (DWs) in layers 1 and 2, which is referred to as the twisted-A double domain wall (TA-2DW) state. On the other hand, when the intralayer exchange energy cost associated with these DWs becomes more important, the system can reduce energy by getting rid of the DW in one or both layers, giving rise to configurations referred to as twisted-A one domain wall (TA-1DW) and collinear (CL) states, respectively. Since the area fraction of the R-stacking and FM (interlayer coupling between layers 2-3) region per moiré cell is larger than that of the M-stacking and AFM region (Supplementary Fig. 11), and the maximum strengths of the interlayer exchange of the FM and AFM regions are approximately the same, FM orientations between layers 2 and 3 are favored over AFM orientations for the CL state. Finally, similar to the bilayer $CrI_3$ case, for weaker perpendicular anisotropy, magnetic configurations with substantial in-plane components (the so-called twisted-S phase[21]) can be favored to lower the interlayer exchange energy via the formation of ~90° DWs (reducing DW widths compared to ~180° DWs). For tDB $CrI_3$, such ~90° DWs can be formed in the middle two layers or all the layers, resulting in configurations labeled as twisted-S two domain wall (TS-2DW) and twisted-S four domain wall (TS-4DW), respectively. By comparing the free energies of the abovementioned configurations (see Supplementary Note 11 for details), we present the phase diagram of ground states for the tDB $CrI_3$ system in Fig. 5b.

## Discussion

Equipped with the phase diagram, we next provide a qualitative understanding of the experimental observations and compare our results with recent observations of moiré magnetism.

We begin by focusing on the twist angle dependence of the MOKE. To this end, we show the trajectory of the corresponding magnetic phases traversed as the twist angle $\theta$ is increased in Fig. 5b. The blue dashed line corresponds to a calculation based on $K$, $A$ and $J_{\perp}^{0}$ values extracted from the untwisted $CrI_3$ system[41, 42]. Since the MOKE signal is proportional to perpendicular magnetization, in Fig. 5c, we also show the calculated net out-of-plane magnetization component ($\overline{m_z}/4$) for different phases. For very small $\theta$ (i.e., large $\alpha$ and $\beta$), the TA-2DW state is the ground state. This is because the small $\theta$ implies the presence of large moiré domains (of size ~$L \propto a/\theta$) with varying signs of interlayer exchange. The system thus prefers to minimize the energy cost within the domains (at the expense of incurring intralayer exchange and anisotropy energy costs in the DWs) by choosing primarily out-of-plane magnetizations with interlayer arrangements following the sign of local interlayer exchange. Similar to the natural four-layer $CrI_3$, the TA-2DW phase carries negligible out-of-plane magnetization (Fig. 5c), consistent with the observed zero MOKE signal around zero twist angle, as shown in Fig. 2. With increasing twist angle $\theta$, the domain wall energy (which increases proportionally to $L$ [11, 21]) becomes comparable to the domain energy (which scales as $L^2$); the system thus transitions to the TA-1DW state, which, crucially, has a nonzero out-of-plane magnetization. Therefore, the observed MOKE signal going from 0 to finite values is consistent with the appearance of the TA-1DW state. As $\theta$ is further increased such that the cost of forming domain walls becomes too high, the system eventually collapses to a collinear phase (CL) with no DWs and vanishing net out-of-plane magnetization. This transition leads to a drop of the MOKE signal back to zero, as observed in Fig. 2.

Based on the above discussion, we attribute the TA-1DW state to be the most likely candidate that gives rise to the emergent FM loops observed experimentally for intermediate twist angles (the anticipated magnetic field evolution of the TA-1DW state under various scenarios is further consistent with our experiments; see Supplementary Figs. 13-16 for details). We remark, however, that lattice relaxation and/or disorder in tDB system can effectively renormalize material parameters. This can give rise to twist-angle dependent trajectories different from the one followed for typical material parameters extracted from the untwisted $CrI_3$ system. To account for this variation, we also show in Fig. 5b red dot-dashed and black dotted lines, which, respectively, correspond to ten times larger and smaller values of the ratio $K/J_{\perp}^{0}$. We notice that, for smaller $K/J_{\perp}^{0}$, the system can traverse through the twisted-S phases, which also carry a nonzero out-of-plane magnetization. However, to identify the presence of such noncollinear phases, other probes sensitive to in-plane magnetization are needed in future research, such as longitudinal MOKE or spatially sensitive spin probes[10, 43].

Using the phase diagram, we also provide a possible explanation for the observed anomalous temperature dependence of the MOKE signal in Fig. 3. In addition to the twist angle, the temperature dependence of the magnetic properties ($K$, $J_{\perp}^{0}$, and $A$) changes the relative strengths of the anisotropy, interlayer and intralayer exchange energies, resulting in a temperature-induced trajectory on the phase diagram. Depending on this trajectory and the starting phase, the overall out-of-plane physical magnetization ($M_z = \overline{m_z} M_s$) can *increase* with temperature before vanishing near the phase transition to the paramagnetic state. Here, $M_s$ is the saturation magnetization. To demonstrate this possibility, we choose the twist angle of $\theta = 0.71°$ (as for the device in Fig. 3), starting from ($\alpha$, $\beta$) = (56, 1230) at low temperature (an exemplary choice of parameters between the red dot-dashed and black dotted lines mentioned above), and track the evolution of the system on the phase diagram as the temperature is increased. To this end, we vary $K$, $J_{\perp}^{0}$, $A$ and $M_s$ according to the power law of temperature ~$(1-T/T_c)^{\gamma}$, where $T_c$ is the Curie temperature, with exponents $\gamma$ that were experimentally obtained to be ~ 2.3, 0.22, 0.22 and 0.125[11, 35, 36]. With increasing temperature (dashed arrow), $\overline{m_z}/4$ indeed shows an initial increase followed by a decrease in the phase diagram. Accordingly, we plot normalized $M_z$ as a function of temperature (inset in Fig. 5c), showing the nonmonotonic temperature dependence, which is qualitatively consistent with our observations in Fig. 3c.

A recent study and preprint from another team[12, 33] also reported noncollinear phases in tDB $CrI_3$; however, the voltage control and detailed ground-state phase diagram have not been presented. While the general existence of noncollinear phases in our study is consistent with these reports, there are a few qualitative and quantitative differences. First, an additional transition is observed at fields larger than 1 T (also seen in 4L $CrI_3$) in the smallest twist-angle (reported to be ~0.5°) tDB $CrI_3$ sample in Ref. 33; our samples do not show such a high-field transition in small-twist-angle samples. More importantly, for all intermediate twist-angle samples, they observe a single abrupt jump in the magnetization when external magnetic fields are applied to reverse the net magnetization of the noncollinear phase ("one-step" reversal); in contrast, we observe two abrupt changes in magnetization ("two-step" reversal). Finally, the critical twist angle up to which noncollinear phases are observed differs quantitatively between the two studies. Such distinct behaviors in different samples could result from the sensitivity of noncollinear phases and their field-dependent evolution to (a) twist angles (given typical accuracy of ~0.5°) and (b) the amount of disorder/material parameters, with possible differences and effects related to lattice reconstruction[10, 44] and inhomogeneity. For example, we performed micromagnetic simulations on the evolution of noncollinear states with magnetic fields and found that the choice of different material parameters can indeed change the switching behavior from a two-step reversal to a one-step reversal (see Supplementary Fig. 14 for a detailed discussion).

To summarize, the main qualitative and robust features explained by our phase diagram include: the emergent noncollinear phase with finite net magnetization at intermediate twist angles (with the TA-1DW state being the most likely candidate responsible for the observations), along with its magnetic field, twist angle and nonmonotonic temperature dependence. The qualitative agreement between the observations and theoretical results supports the presence of noncollinear phases in tDB $CrI_3$. We note, however, that quantitative differences exist, for example, in the critical twist angles for the transition between the phases (also in comparison with similar work in twisted $CrI_3$ [12, 33]) (see Supplementary Fig. 12) and the functional forms of anomalous temperature dependence of MOKE. These differences could arise from the sensitivity of the phase diagram to material parameters, the presence of disorder[10, 34], or the trapping of experimentally observed spin configurations into metastable states.

## Conclusions

We have reported electrically tunable moiré magnetism in twisted double bilayer $CrI_3$, with emergent magnetic orders that can be interpreted by a phase diagram associated with the moiré wavelength and magnetic parameters of the sample. Our findings suggest that voltage is an effective approach to control magnetic orders and magnetic switching, suggesting that the system could have potential applications in memory and spin-logic devices. The nontrivial magnetic phases could also potentially host topological skyrmion lattices[24] and magnon networks[23], and could be probed using spatially resolved measurements. As a novel degree of freedom, the twist angle is applicable to a range of vdW homo/heterobilayer magnets, including ferromagnets, antiferromagnets, multiferroics, and even quantum spin liquid candidates[14, 45], opening the opportunity to pursue new physics as well as spintronic applications.

## Methods

**Crystal growth.** Single crystal $CrI_3$ was synthesized using the chemical vapor transport (CVT) method[46]. The Cr powder and iodine pieces were mixed with a stoichiometric ratio and loaded into a quartz tube (inner diameter, 10 mm; length, 180 mm). The quartz tube was sealed under vacuum and then transferred to a double-temperature-zone furnace. The temperatures of the hot and cold ends of the furnace were set at 650 °C and 550 °C, respectively. The growth with such a temperature gradient lasted for 7 days. Finally, the furnace was shut down, and the quartz tube was naturally cooled down to room temperature. The black plate-like $CrI_3$ crystals can be found at the cold end of the quartz tube.

**Fabrication of twisted $CrI_3$ and gated devices.** Flakes of $CrI_3$ were obtained by exfoliation of bulk material onto a silicon wafer with 285 nm oxide. The bilayers were selected by optical contrast and later confirmed by AFM and MOKE measurements. The tear-and-stack technique[13, 28] was employed in this work to fabricate twisted double bilayers. We first used a polydimethylsiloxane/polycarbonate stamp to pick up an hBN flake. Then, we carefully controlled the hBN flake to contact one part of the selected bilayer $CrI_3$ flake. After lifting up the stamp, the contacted part of the $CrI_3$ flake was torn off by the hBN layer, and the remaining part was left on the silicon wafer. Then, the remaining part on the wafer was rotated with a target twist angle $\theta$, aligned with the separated part on the hBN flake, and picked up to form the twisted stack. Afterwards, another hBN flake was picked up, so the stacked $CrI_3$ layers were sandwiched between hBN and protected from degradation. Typical $CrI_3$ flakes before and after stacking are shown in Fig. 2a. To fabricate the back-gated device, few-layer graphene flakes were exfoliated and picked up during the stacking processes as the contact to tDB $CrI_3$. The stack was dropped onto prepatterned gold electrodes on a silicon wafer. The silicon oxide and bottom hBN flake act as back-gate dielectric layers. All exfoliation and stacking operations were performed inside an argon glovebox to avoid degradation. The exposure time to air was kept below ten minutes while transferring the fabricated sample into the measurement chamber before pumping down.

**TEM characterization.** To fabricate TEM samples, very thin (<5 nm) top and bottom hBN flakes were used to improve the dark-field image contrast. The Holey silicon nitride support membrane can suspend the hBN/tDB $CrI_3$/hBN stack for TEM imaging. The twisted sample was characterized by scanning TEM equipped with a high-angle annular dark-field (HAADF) detector on a ThermoFisher Scientific Themis Z Aberration-corrected Transmission Electron Microscope. The instrument was operated at 300 kV and 0.25 nA current.

**Polar-MOKE microscopy.** The polarization of linearly polarized light reflected from a magnetic material can be rotated by a Kerr angle $\theta_K$, which is proportional to the magnetization of the material. In this work, the incident light is normal to the sample plane, and the MOKE is in the polar geometry, meaning that the magnetic vector being probed is perpendicular to the sample surface and parallel to the incident light. A balanced photodetector and lock-in method are used to obtain the MOKE signal[47]. The incident laser used here has a wavelength of 633 nm and a power of 5 μW. The sample is placed in a helium-flow optical cryostat with a base temperature of 6 K and magnetic field (perpendicular to the sample surface) up to 5 T. The laser is focused onto the sample surface by an objective with a spot diameter of 0.5 μm. Note that in principle, in-plane components of spins can emerge in twisted magnets even with perpendicular magnetic anisotropy[21]. A perpendicular field can induce canting of the in-plane spins (if any), contributing to a continuously varying MOKE background[48], which is typically subtracted and eliminated from our MOKE signal (Supplementary Fig. 17). Therefore, we mainly focus on the out-of-plane components of spins with spin-flip transitions in this work.

## Data availability

Data are available in the manuscript or supplementary materials. Additional data are available from the authors upon reasonable request.

## Code availability

The computer code used in this study is available from the corresponding authors upon reasonable request.

**Acknowledgements**

We acknowledge partial support of the work from US Department of Energy (DOE) Office of Science through the Quantum Science Center (QSC, a National Quantum Information Science Research Center) and Department of Defense (DOD) Multidisciplinary University Research Initiatives (MURI) program (FA9550-20-1-0322) for materials and device fabrication and MOKE and TEM measurements. G.H.C. and Y.P.C. also acknowledges partial support from WPI-AIMR, JSPS KAKENHI Basic Science A (18H03858), New Science (18H04473 and 20H04623), and Tohoku University FRiDUO program in early stages of the research. M.R., A.R. and P.U. also acknowledge the support from the National Science Foundation (NSF) (ECCS-1810494). Z.Q.M. acknowledges the support by the US DOE under grant DE-SC0019068 for sample synthesis. K.W. and T.T. acknowledge support from the Elemental Strategy Initiative conducted by the MEXT, Japan (JPMXP0112101001) and JSPS KAKENHI (19H05790, 20H00354 and 21H05233).

**Author contributions**

G.H.C. and Y.P.C. conceived the project. G.H.C. fabricated the devices and performed experiments, assisted by A.L.A. M.M.R., A.R. and P.U. performed supporting theoretical analysis. X.T.L., G.H.C., L.L. and L.F. performed the TEM measurements. Y.L.Z. and Z.Q.M. provided bulk $CrI_3$ crystals. K.W. and T.T. provided bulk hBN crystals. Y.P.C. and P.U. supervised the project. G.H.C., M.M.R., P.U. and Y.P.C. wrote the manuscript with input from all co-authors.

**Competing interests**

The authors declare no competing financial interests.

**Figure Captions**

**Fig. 1 | Moiré superlattice and STEM characterizations of tDB $CrI_3$. a**, The moiré superlattice structure of tDB $CrI_3$ with a twist angle $\theta$ between the two bilayers. Only Cr atoms of the middle two $CrI_3$ layers are shown for simplicity with red and blue balls belonging to each of the two layers. Regions of AA stacking, monoclinic (M) stacking, and rhombohedral (R) stacking are indicated by blue, green and red circles, respectively. The rhombus-shaped lines indicate the moiré primitive cell with a period of $L$. Right: The middle two layers of tDB $CrI_3$ are twisted, and the top/bottom $CrI_3$ bilayers (2L) retain monoclinic stacking. Bottom: Noncollinear spin textures of the monoclinic-rhombohedral domain wall formed in the twisted middle two layers. **b,** FFT pattern of the STEM image (Supplementary Fig. 2) of a tDB $CrI_3$ with a target twist angle of 1.42°. Green and red rectangles mark the first- and second-order Bragg peaks, respectively. Right: The magnification of one peak shows two slightly separated peaks from the top and bottom bilayer $CrI_3$ with a measured twist angle of ~1.9°(±0.3°). **c,d,** Real-space moiré patterns reconstructed by inverse FFT of the two sets of Bragg peaks marked by green and red rectangles in **b**. Rhombus-shaped lines indicate primitive cells with periods related to moiré wavelength $L$.

**Fig. 2 | Twist angle dependence of the magnetic behaviors in tDB $CrI_3$. a**, Optical micrograph of a bilayer $CrI_3$ flake (left) and a 5°-tDB $CrI_3$ (right) made from it by the tear-and-stack technique. The top and bottom bilayers of $CrI_3$ are outlined by the blue and red dashed lines, respectively. Thicker flakes are outlined by gray dashed lines. Scale bars are 5 µm. **b,c,** MOKE signal as a function of perpendicular magnetic field in bilayer $CrI_3$ and 5°-tDB $CrI_3$ (both from the same $CrI_3$ flake), measured at the yellow and red spots in **a**, respectively. Magnitudes of the FM loop and AFM spin-flip transition are denoted by $\Delta\theta_{K1}$ and $\Delta\theta_{K2}$. Insets in **b** depict the magnetic ground states of bilayer $CrI_3$. The potential spin orientations in each layer for **c** are shown in Fig. S16. **d,e,** MOKE signal as a function of magnetic field in 0°- and 30°-tDB $CrI_3$. **f,** Fraction of MOKE magnitudes $\Delta\theta_{K1}/(\Delta\theta_{K1}+\Delta\theta_{K2})$ as a function of the twist angle ($\theta$, in log scale). The data points at each angle are measured at different positions of each sample. Data (zero value) at $\theta$ of 0°, 30°, and 45° each have at least two reproducible points overlapping. Blue-, pink- and cyan-shaded areas are guides to the eyes. The twist angle accuracy is ~0.5°. The error bars are the uncertainties in extracting the MOKE magnitudes, including the difference between forward and backward sweeps of the magnetic field and the standard deviation ($n$ = 100).

**Fig. 3 | Temperature dependence of the magnetic behaviors in tDB $CrI_3$. a,b,** MOKE signal as a function of magnetic field at representative temperatures in bilayer $CrI_3$ and 0.71°-tDB $CrI_3$ (from the same bilayer $CrI_3$ flake), respectively. Curves are vertically shifted for clarity. The data at 48 K are scaled by a factor of 4 for better comparison. **c,** MOKE magnitudes of the bilayer AFM spin-flip transition $\Delta\theta_{K0}$, the tDB FM loop $\Delta\theta_{K1}$, the AFM spin-flip transition $\Delta\theta_{K2}$ and the total magnitude $\Delta\theta_{K1} + \Delta\theta_{K2}$ as functions of temperature. All these values are extracted from MOKE results in bilayer $CrI_3$ sample and tDB $CrI_3$ sample, as shown in **a** and **b**. Solid lines are phenomenological fits with the power-law form, as the guide to the eyes. The error bars are the uncertainties in extracting the MOKE magnitudes, including the difference between forward and backward sweeps of the magnetic field and the standard deviation ($n$ = 100).

**Fig. 4 | Electrical control of the magnetic behaviors in tDB $CrI_3$. a,b,** Optical micrograph ($\theta \sim 0.14°$) and schematic side view of a tDB $CrI_3$ device with a back-gate voltage $V_{bg}$ applied. A few-layer graphene (FLG) flake is used as the contact to the stack. Scale bar is 5 μm. **c,** MOKE signal as a function of magnetic field at a gate voltage of -80 V. Field sweeping directions are denoted by the arrows. Four plateaus at low fields are denoted by A, B, C and D. **d,** MOKE signal as functions of magnetic field and back-gate voltage $V_{bg}$. The top and bottom panels correspond to backward and forward sweeps of the magnetic field, respectively. The sharp color boundaries marked by arrows indicate the FM coercivity and AFM spin-flip transitions. **e,** Gate-voltage-controlled magnetic switching. Voltage sweeping directions are denoted by the arrows. The sample is initialized by a high magnetic field of +(-)1 T and then biased at fields of ① -0.141 T, ② -0.151 T, ③ -0.16 T, (④ 0.14 T, ⑤ 0.15 T, ⑥ 0.153 T), respectively, corresponding to the circled numbers in **c**. The intermediate states (minor plateaus) seen in **c** and **e** during A-to-B and C-to-D switching may be due to domain wall motion with pinning effect by defects[34, 36].

**Fig. 5 | Theoretical analysis of the moiré magnetism in tDB $CrI_3$. a,** Schematics of the magnetic phases: twisted-A phase with double domain walls (TA-2DW), twisted-A phase with a single domain wall (TA-1DW), twisted-S phase with double domain walls (TS-2DW), twisted-S phase with four domain walls (TS-4DW), and collinear phase (CL, no domain walls). Layers 1-4 are labeled from bottom to top. Monoclinic (M) and rhombohedral (R) stacking are labeled for each layer pair (sandwiching the label), and M/R regions between the middle two layers are indicated by green/pink colors. Out-of-plane (in-plane) spins in the four layers are denoted by vertical (horizontal) arrows. The red crosses denote the magnetic domain walls where noncollinear spin textures exist. **b,** Magnetic phase diagram showing the emergence of noncollinear ground states as functions of dimensionless parameters $\alpha$ and $\beta$. Colored dots denote the simulated cases. The red dot-dashed, blue dashed, and black dotted lines correspond to constant $K/J_{\perp}^{0}$ ratios of 72.3, 7.23, and 0.723, respectively, suggesting phase transitions with the twist angle. **c,** Calculated net out-of-plane component of the dimensionless magnetization ($\overline{m_z}/4$) normalized by the value for fully spin-polarized states (i.e., out-of-plane spins ↑↑↑↑ in all layers gives $\overline{m_z}/4 = 1$) as functions of $\alpha$ and $\beta$. The dashed arrow is an exemplary trajectory of increasing temperature (T, starting from 7 K, with corresponding ($\alpha, \beta$) = (56, 1230) marked by the black dots in both **b** and **c**), giving a sharp drop in $\beta$ and a nonmonotonic behavior of the overall out-of-plane physical magnetization $M_z$ shown in the inset, for $\theta = 0.71°$.

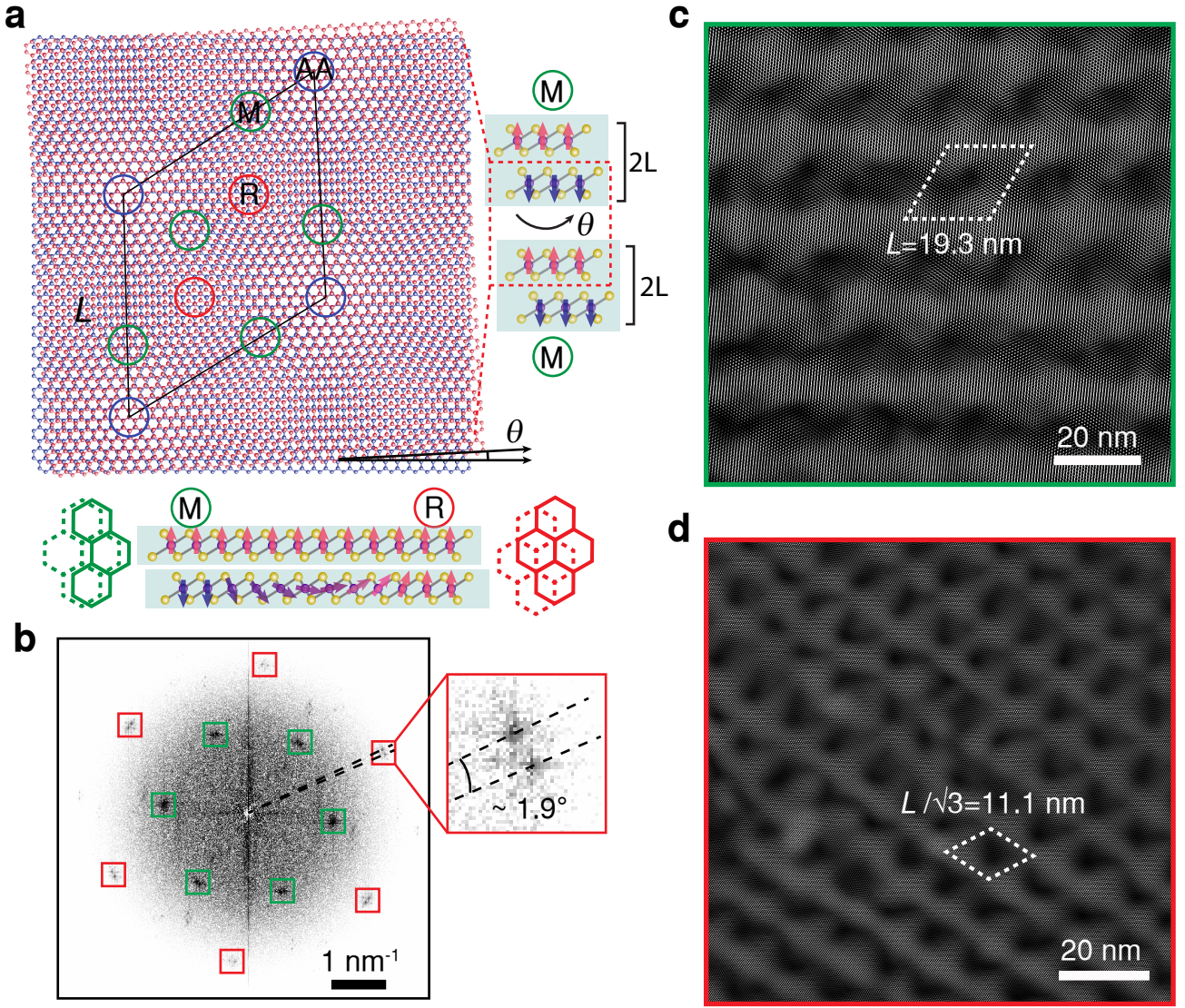

a

AA
M
R
L
M
2L
θ
2L
M
θ
M
R

c

L=19.3 nm
20 nm

b

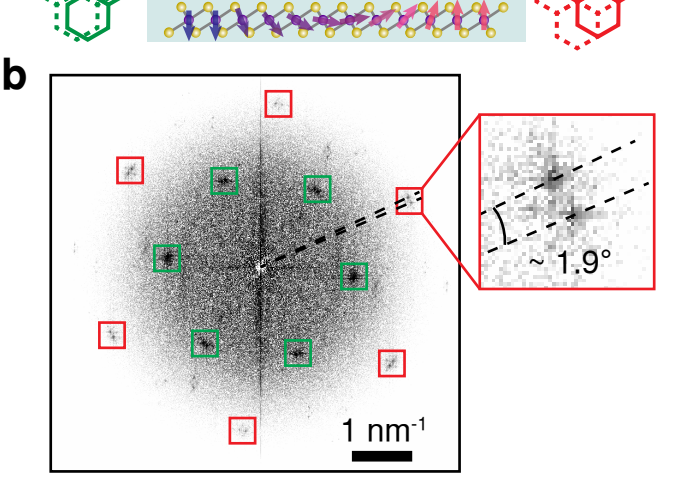

d

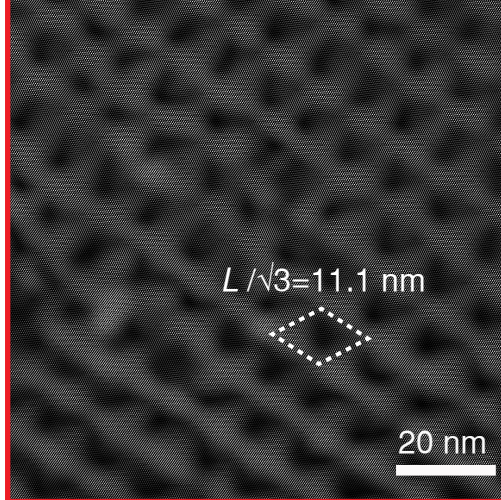

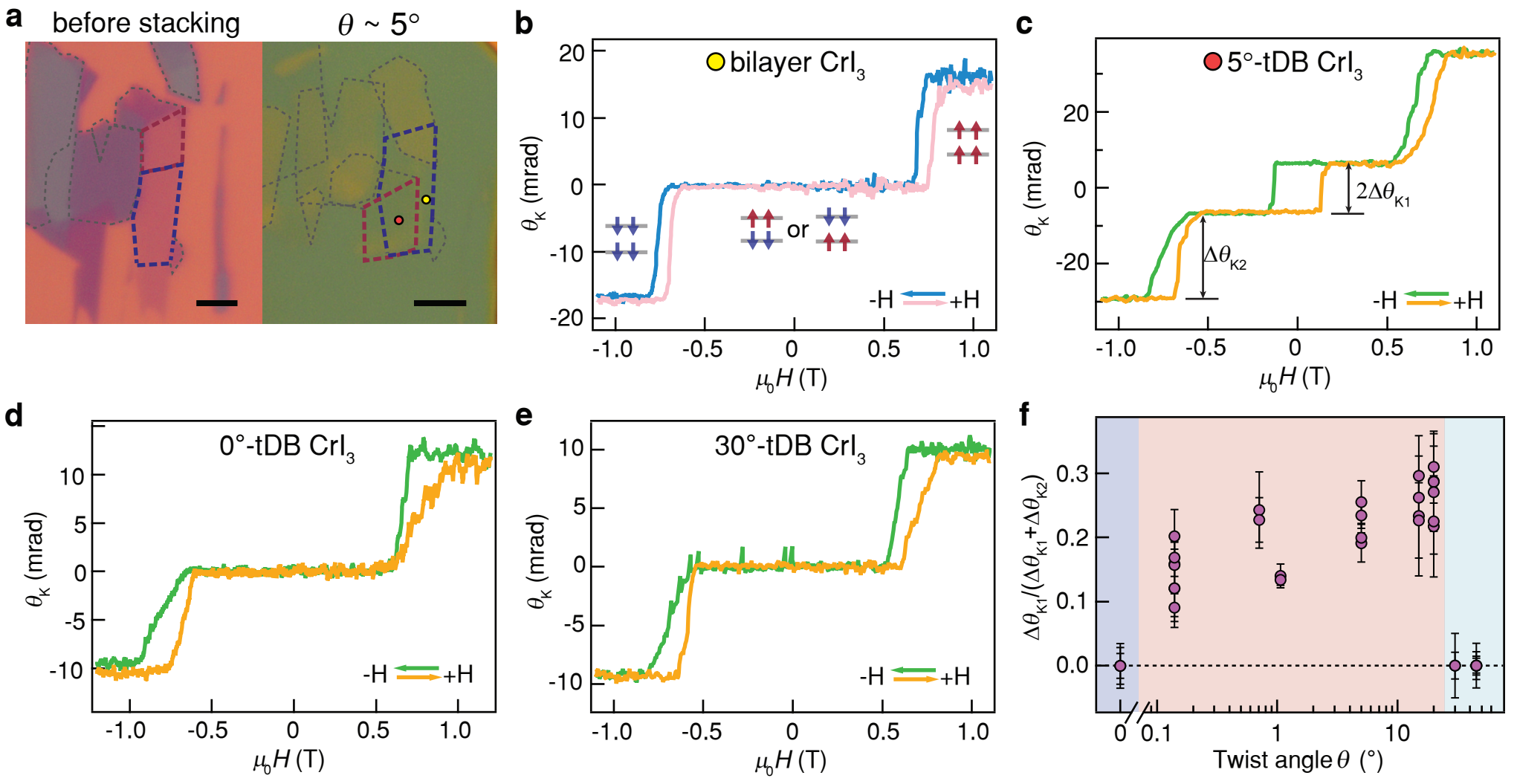

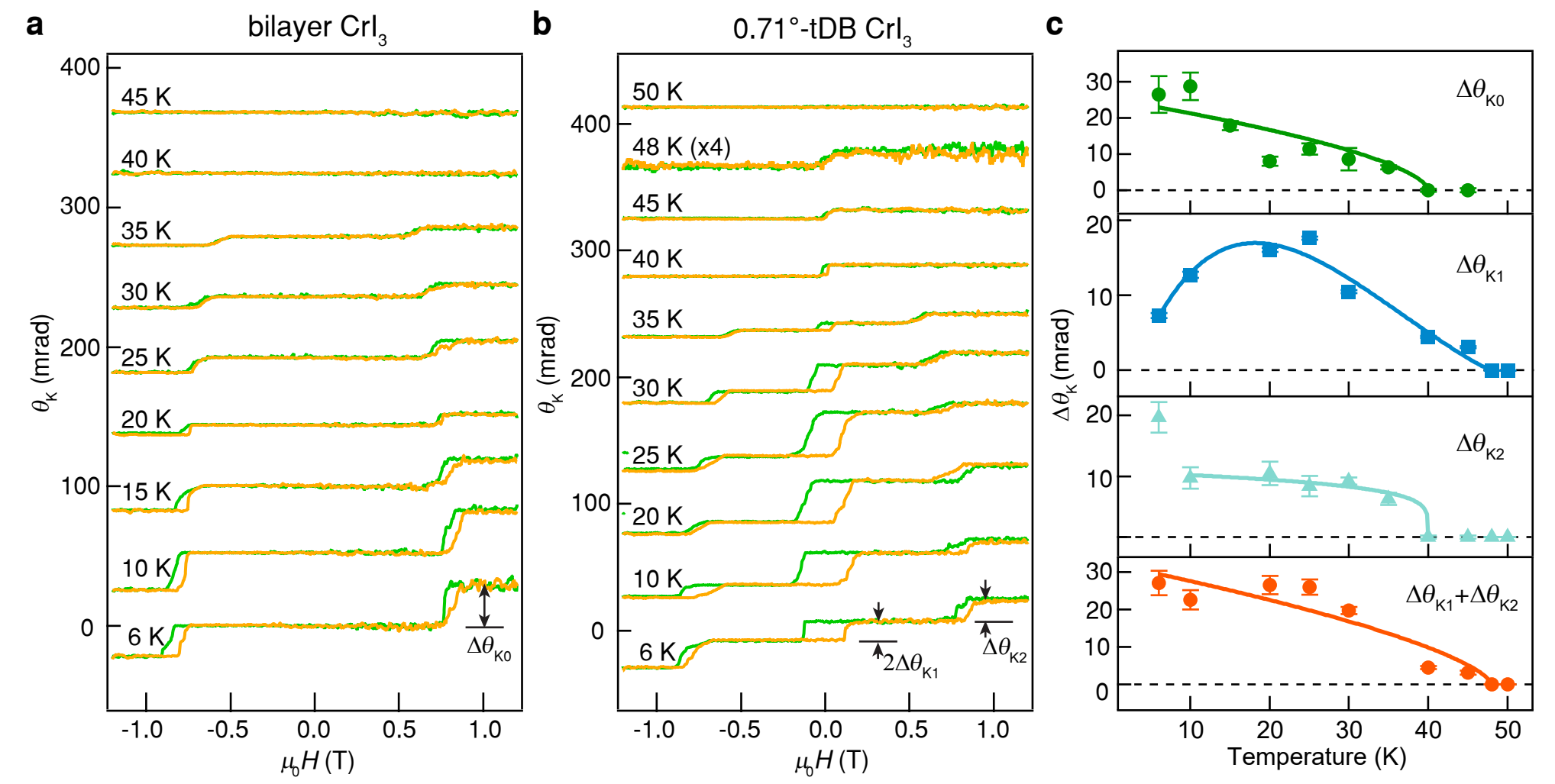

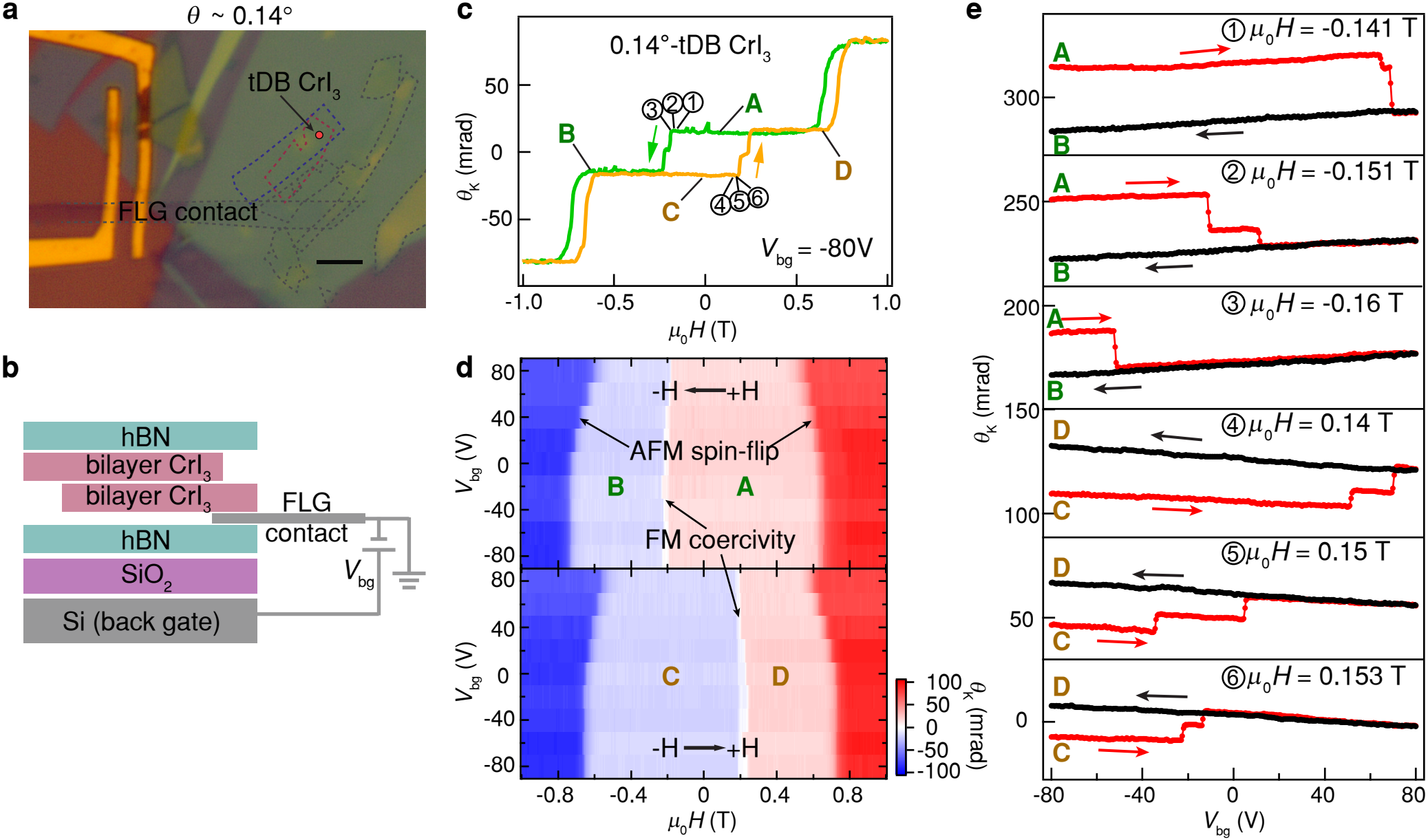

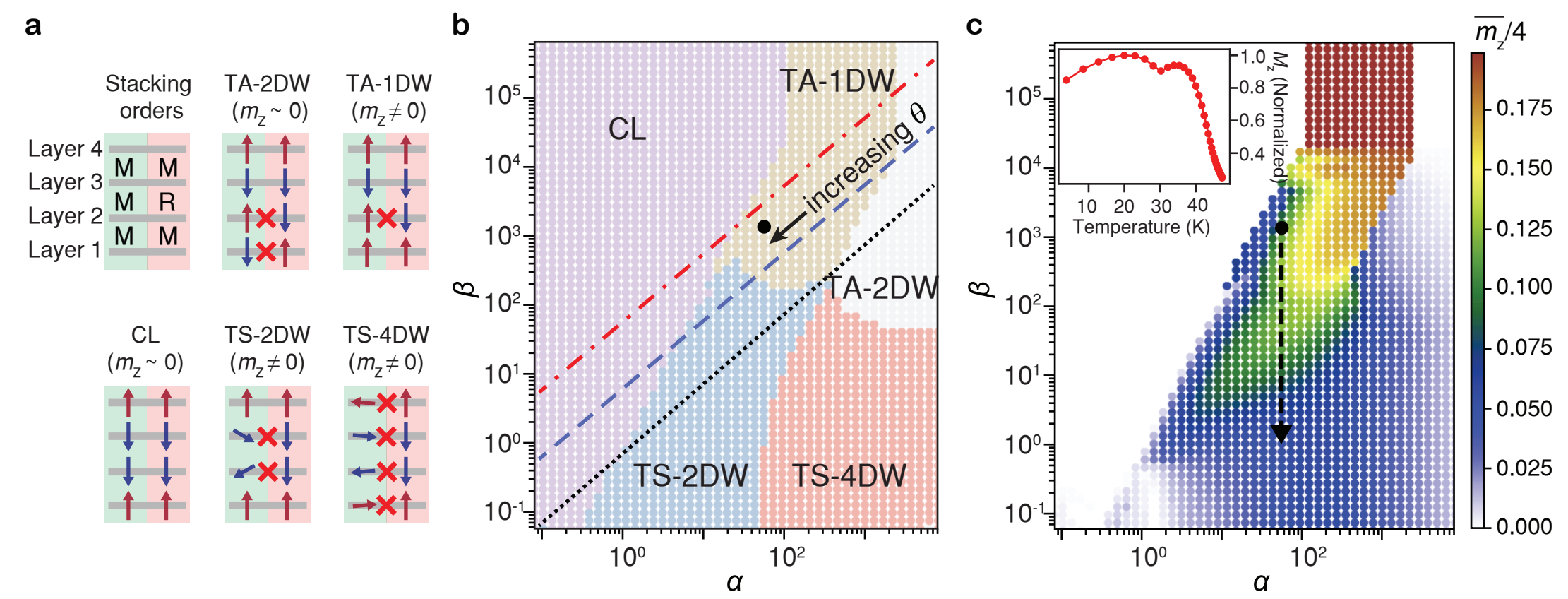